# RADIATION FROM A CHARGED PARTICLE FLYING THROUGH A DIELECTRIC BALL

Svetlana R. ARZUMANYAN

*Institute of Applied Problems in Physics, 25 Hr. Nersessian Street, 0014Yerevan, Armenia*

**Abstract**
The radiation from a relativistic charged particle uniformly moving through the centre of a dielectric ball is investigated. Analytical expressions for spectral and spectral-angular distributions of the radiated energy are derived and numerical results for the spectrum of radiation are given. It is shown, that the spectral distribution of Cherenkov radiation generated by the relativistic particle inside a dielectric ball at specified frequencies is strongly influenced by the ball-vacuum boundary.

## 1. Introduction

The presence of matter may essentially influence the characteristics of high energy electromagnetic processes, generating new types of phenomena. The operation of a number of devices intended for production of electromagnetic radiation is based on the interaction of relativistic electrons with matter [1].

The synchrotron radiation from a charged particle rotating in a homogeneous medium is considered in [2] (see also [3,4]). It has been shown that the interference between synchrotron radiation and Cherenkov radiation (CR) leads to interesting consequences. New interesting phenomena occur in the case of inhomogeneous media.

The investigations of radiation from a charged particle rotating along the equatorial orbit about/inside a dielectric ball showed [5,6] that when the Cherenkov condition for the ball material and particle speed is satisfied, there appear high narrow peaks in the spectral distribution of the number of electromagnetic field quanta emitted to the outer space for some specific values of the ratio of ball-to-particle orbit radii. In the vicinity of these peaks the radiated energy exceeds the corresponding value for the case of a homogeneous and unbounded medium by several orders of magnitude. The absorption and dispersion of the electromagnetic waves inside the ball material were taken into account in [7]. It is worthwhile to point that the similar phenomenon takes place in the case of cylindrical symmetry for e.g. the radiation emitted from a charged particle rotating about/inside a dielectric cylinder [8] ( for a more general case of a particle moving along a helical orbit see [9]).

The radiation from a relativistic particle moving along a rectilinear trajectory passing through the centre of a dielectric or metallic ball was investigated in [10] (see also [11]). However, in [10,11] the phenomena of the absorption and dispersion of electromagnetic waves inside the ball material were not considered. In the present paper the dielectric losses of electromagnetic waves inside the ball material are taken into account, and the features of CR from a relativistic particle conditioned by the influence of the ball surface are analyzed.

The paper is organized as follows. In Section 2 the description of the problem and the analytical expressions for the spectral and spectral-angular distributions of the radiation energy are given. In Section 3 numerical results are presented for the spectrum of the radiation from a relativistic electron passing through a ball made of melted quartz. The main results are summarized in the last section.

## 2. Formulation of problem and basic formulas

In [12,13], a method is proposed for the evaluation of the Green function of the electromagnetic field in a medium consisting of $N \geq 1$ spherically-symmetric layers having a common center and different permittivities. Based on this method, a formula is derived for the energy of radiation from a charged particle moving along an arbitrary trajectory in such medium. The special cases of a charged particle rotating about or inside a dielectric ball have been investigated in [5-7]. In the present paper another example is investigated when a charged particle flies rectilinearly and uniformly in a medium with the permittivity $\varepsilon_1$ and passes through a centre of a dielectric ball with permittivity $\varepsilon_0$ immersed in such medium (see Fig. 1). The magnetic permeability of matter is taken to be 1.

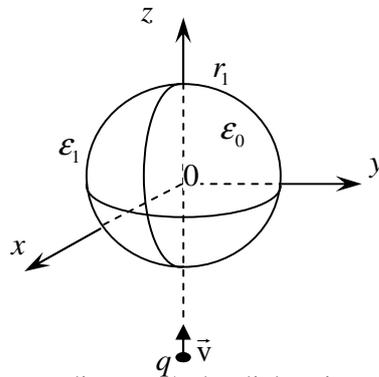

Fig. 1. A particle flying from a medium with the dielectric permittivity $\varepsilon_1$ into a ball with the radius $r_1$ and the dielectric permittivity $\varepsilon_0$. The trajectory of the particle is described by the equation $z = \mathrm{v}t$.

In the case under consideration the current density can be written as

$$\vec{j}(\vec{r},t) = \frac{q\vec{v}}{\pi r^2 \sin\theta}\begin{cases} \delta(r-vt)\delta(\theta) & \text{for} \quad t>0, \\ \delta(r+vt)\delta(\theta-\pi) & \text{for} \quad t<0, \end{cases} \quad (1)$$

where $q$ and $v$ are the charge and velocity of the particle, and $\theta$ is the polar angle of corresponding spherical system of coordinates $r,\theta,\varphi$. For the corresponding Fourier transform one has

$$\vec{j}(\vec{r},\omega) = \frac{1}{2\pi}\int \vec{j}(\vec{r},t)e^{i\omega t}dt = \frac{q\vec{v}}{2\pi^2 vr^2 \sin\theta}[\delta(\theta)e^{i\frac{\omega}{v}r} + \delta(\theta-\pi)e^{-i\frac{\omega}{v}r}]. \quad (2)$$

Expanding $\vec{j}(\vec{r},\omega)$ in spherical vectors $\vec{X}_{lm}^{(\mu)}$ [14], we can evaluate the corresponding expansion coefficients:

$$j_\mu^{lm}(r) \equiv \int \vec{j}(\vec{r},\omega)\vec{X}_{lm}^{(\mu)*}d\Omega = \frac{q\delta_{m0}}{\pi v r^2}[e^{i\frac{\omega}{v}r}\int \delta(\theta)\vec{v}\vec{X}_{l0}^{(\mu)}d\theta + e^{-i\frac{\omega}{v}r}\int \delta(\theta-\pi)\vec{v}\vec{X}_{l0}^{(\mu)}d\theta]. \quad (3)$$

As a result we find

$$j_1^{lm} = \frac{q\delta_{m0}}{2\pi r^2}\sqrt{\frac{2l+1}{4\pi}}[e^{i\frac{\omega}{v}r} - (-1)^l e^{-i\frac{\omega}{v}r}], \quad j_2^{lm} = j_3^{lm} = 0. \quad (4)$$

Similarly, one can expand the vector potential $\vec{A}(\vec{r},\omega)$ of the electromagnetic field in spherical vectors $\vec{X}_{lm}^{(\mu)}$. In the Lorentz gauge the expansion coefficients $A_\mu^{lm}$ are given by the expressions (for details see [13])

$$(2l+1)\frac{cr_1^2}{4\pi}\begin{bmatrix}A_1^{lm}(r)\\A_2^{lm}(r)\end{bmatrix} = \begin{bmatrix}lu_{l-1}+(l+1)u_{l+1}\\\sqrt{l(l+1)}(u_{l-1}-u_{l+1})\end{bmatrix}_r^{lm} + $$
$$+ \gamma_l[lB_{l-1}^{lm} - (l+1)B_{l+1}^{lm}]\begin{bmatrix}lP_{l-1}+(l+1)P_{l+1}\\\sqrt{l(l+1)}(P_{l-1}-P_{l+1})\end{bmatrix}_{(r,r_1)}, \quad (5)$$

The magnetic type multipoles ($\mu=3$) are not generated: $A_3^{lm}(r) = 0$. Here

$$u_{\underline{l}}^{lm}(r) \equiv \int_0^\infty P_{\underline{l}}(r,x)j_1^{lm}(x)x^2 dx = \frac{q\delta_{m0}}{2\pi}\sqrt{\frac{2l+1}{4\pi}}\int_0^\infty P_{\underline{l}}(r,x)[e^{i\frac{\omega}{v}x} - (-1)^l e^{-i\frac{\omega}{v}x}]dx,$$

$$B_{\underline{l}}^{lm} = \frac{q\delta_{m0}}{2\pi}\sqrt{\frac{2l+1}{4\pi}}\{\lambda_0 j_l(\lambda_0 r_1)\int_{r_1}^\infty h_{\underline{l}}^{(1)}(\lambda_1 x)[e^{i\frac{\omega}{v}x} - (-1)^l e^{-i\frac{\omega}{v}x}]dx +$$
$$+ \lambda_1 h_l^{(1)}(\lambda_1 r_1)\int_0^{r_1} j_{\underline{l}}(\lambda_0 x)[e^{i\frac{\omega}{v}x} - (-1)^l e^{-i\frac{\omega}{v}x}]dx\}, \quad (6)$$

with the notations

$$\gamma_l = \frac{1/\varepsilon_0 - 1/\varepsilon_1}{lz_{l-1}^l + (l+1)z_{l+1}^l}, \qquad z_v^l = \frac{\lambda_1 j_v(\lambda_0 r_1) h_l^{(1)}(\lambda_1 r_1)/\varepsilon_1 - \lambda_0 j_l(\lambda_0 r_1) h_v^{(1)}(\lambda_1 r_1)/\varepsilon_0}{\lambda_1 j_v(\lambda_0 r_1) h_l^{(1)}(\lambda_1 r_1) - \lambda_0 j_l(\lambda_0 r_1) h_v^{(1)}(\lambda_1 r_1)}, \qquad (7)$$

and $h_l^{(1)}(y) = j_l(y) + in_l(y)$ ($j_l(y)$ and $n_l(y)$ are the spherical Bessel and Neumann functions respectively). In (6),(7) the following notations are used

$$f_l(\tau) \equiv f_l(\tau)/a_l^{(12)}, \quad \|a_l^{ik}\| = \begin{bmatrix} [h_l^{(1)}(\lambda_0 r_1), h_l^{(1)}(\lambda_1 r_1)] & -[j_l(\lambda_0 r_1), h_l^{(1)}(\lambda_1 r_1)] \\ -[j_l(\lambda_0 r_1), h_l^{(1)}(\lambda_1 r_1)] & [j_l(\lambda_0 r_1), j_l(\lambda_1 r_1)] \end{bmatrix}, \qquad (8)$$

$$[a_l(\lambda_0 r_1), b_l(\lambda_1 r_1)] \equiv \lambda_1 a_l(\lambda_0 r_1) b_{l-1}(\lambda_1 r_1) - \lambda_0 a_{l-1}(\lambda_0 r_1) b_l(\lambda_1 r_1),$$

and the definition of the function $P_l(r,x)$ is given in [13]. For the case under consideration, $r > r_1$, one obtains

$$P_l(r, r_1) = h_l^{(1)}(\lambda_1 r) j_l(\lambda_0 r_1). \qquad (9)$$

From the relations

$$h_l^{(1)}(x) \approx (-i)^{l+1} \frac{e^{ix}}{x}, \qquad j_l(x) \approx \frac{1}{x}\sin(x - l\pi/2), \qquad x \gg l, \qquad (10)$$

for the spherical Bessel and Hankel functions [15,16] it follows that at large distances from the ball one has

$$A_\mu^{lm}(r) \approx \delta_{m0} a_\mu^l \frac{e^{i\lambda_1 r}}{\lambda_1 r} + \ldots \qquad r \to \infty \qquad (11)$$

Here

$$\begin{bmatrix} a_1^l \\ a_2^l \end{bmatrix} = \frac{(-i)^l q}{cr_1^2 \sqrt{\pi(2l+1)}} \begin{bmatrix} l w_{l-1} - (l+1) w_{l+1} \\ \sqrt{l(l+1)}(w_{l-1} + w_{l+1}) \end{bmatrix}, \qquad a_3^l = 0 \qquad (12)$$

and

$$w_{l_1} \equiv J_{l_1}(<) + i\lambda_1 r_1^2 \left[ a_l^{(12)} J_{l_1}(>) + a_l^{(22)} H_{l_1} \right] + \gamma_l j_{l_1}(\lambda_0 r_1)\left[ l Q_{l-1} - (l+1) Q_{l+1} \right]. \qquad (13)$$

In (13) the following notations are introduced

$$J_{l_1}(<) \equiv \int_0^{r_1} j_{l_1}(\lambda_0 x)[e^{i\frac{\omega}{v}x} - (-1)^l e^{-i\frac{\omega}{v}x}]dx, \qquad J_{l_1}(>) \equiv \int_{r_1}^\infty j_{l_1}(\lambda_1 x)[e^{i\frac{\omega}{v}x} - (-1)^l e^{-i\frac{\omega}{v}x}]dx,$$

$$H_{l_1} \equiv \int_{r_1}^\infty h_{l_1}^{(1)}(\lambda_1 x)[e^{i\frac{\omega}{v}x} - (-1)^l e^{-i\frac{\omega}{v}x}]dx, \qquad Q_{l_1} = \lambda_0 j_l(\lambda_0 r_1) H_{l_1} + \lambda_1 h_l^{(1)}(\lambda_1 r_1) J_{l_1}(<). \qquad (14)$$

According to [12] the spectral-angular and spectral distributions of the radiation energy (during the all time of charge motion) are determined by the expressions

$$\frac{dI}{d\omega d\Omega} = \frac{c}{\sqrt{\varepsilon_1}} \left| \sum_l a_2^l \vec{X}_{l0}^{(2)} \right|^2 \qquad (15)$$

and

$$\frac{dI}{d\omega} = \frac{c}{\sqrt{\varepsilon_1}} \sum_l \left| a_2^l \right|^2. \qquad (16)$$

respectively.

## 3. Numerical results

The dependence of dimensionless function $dI/\hbar d\omega$ on the cyclic frequency $\omega$ of the radiation is given in Fig.2 for an electron flying through a dielectric ball surrounded by the vacuum ($\varepsilon_1 = 1$). The radius of the ball and its permittivity are equal $r_1 = 4\,\text{cm}$ and $\varepsilon_0 = \varepsilon_0' + i\varepsilon_0'' = 3.78(1 + 0.0001i)$ [15] respectively (dispersion of electromagnetic waves inside the ball made of melted quartz is not taken into account). We consider the radiation with the wavelength of the order of ball size ($10 \leq \omega \leq 70\,\text{GHz}$) and the electron with energy 2MeV, so that the Cherenkov condition

$$v > c/\sqrt{\varepsilon_0'} \qquad (17)$$

for the velocity of the electron and the substance of the ball is satisfied.

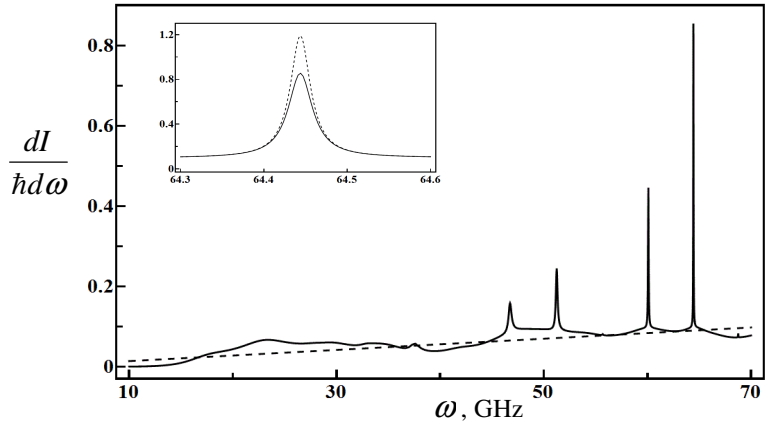

Fig. 2. Spectral density of the radiation from an electron with the energy 2MeV passing through the centre of a ball with permittivity $\varepsilon_0 = \varepsilon_0' + i\varepsilon_0'' = 3.78(1 + 0.0001i)$ and radius $r_1 = 4\,\text{cm}$ (continuous curve). The dotted curve corresponds to the case when the same electron flies in the infinite and continuous medium with $\varepsilon_0 = \varepsilon_0'$ and the radiation is accumulated from the path length equal to $2r_1$ (the diameter of ball). The Cherenkov condition (17) is satisfied.

The dashed line in Fig. 2 corresponds to CR from an electron with the same energy moving in an infinite transparent medium with $\varepsilon_0 = \varepsilon_0'$ under the condition that the radiation is collected from the path having the length $2r_1$ (diameter of the ball). The radiation intensities (full and dashed lines) considerably differ only in the vicinity of the dedicated "resonant" frequencies (for example, $\omega \approx 64.45$ GHz) with the wavelengths of the order of $r_1$. At these frequencies the values of $dI/\hbar d\omega$ are almost ten times larger than those for neighboring frequencies. The presence of the peaks indicates that the spectrum of CR generated by the particle inside the substance of the ball is strongly modified due to the influence of the ball-vacuum boundary. Moreover, if the dielectric losses of the radiated energy inside the substance of the ball are small, then the quantity $dI/\hbar d\omega$ practically is not influenced outside the regions near the peaks, and it considerably increases at "resonant" frequencies $\omega \approx \omega_r$. This conclusion follows from the comparison of the curves in the top left corner of Fig.2, where the dotted and continuous curves correspond to the ball made of loss-free dielectric and of melted quartz respectively. One can explain the abovementioned increase of the radiation intensity as result of the constructive superposition of the electromagnetic waves numerously reflected from the internal surface of the ball. Note that the situation here is qualitatively different compared to the case of dielectric plate. CR from the electron with energy 2MeV is completely reflected from the boundaries of plate with dielectric permittivity $\varepsilon_0 = 3.78(1 + 0.0001i)$ immersed in the vacuum and it is confined inside the plate. Only the transition radiation penetrates outside the plate. But the intensity of transition radiation is considerably smaller than that generated inside the ball and depicted in Fig.2.

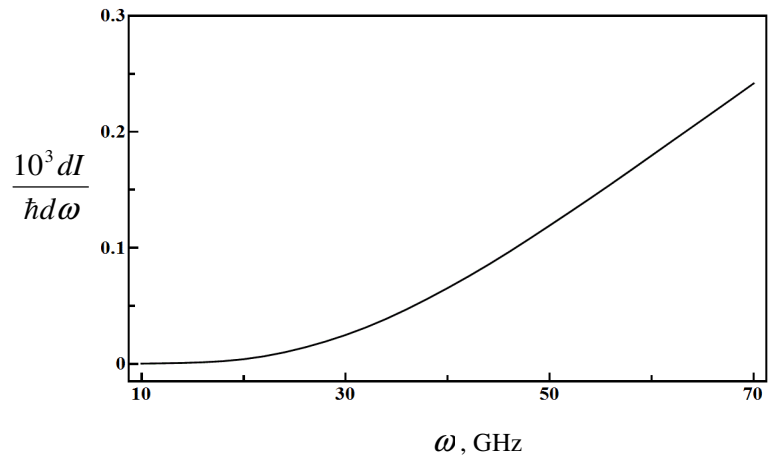

Fig. 3. Spectral distribution of the radiation energy from an electron passing through the centre of the ball when the Cherenkov condition (17) is not satisfied (the dielectric constant of the ball $\varepsilon_0 = 1.05$, the values of the other parameters are the same as those in Fig. 2).

The spectral distribution of radiated energy is plotted in Fig.3 for the case when the Cherenkov condition (17) is not satisfied ($\varepsilon_0 = 1.05$, the values of the other parameters are the same as those in Fig.2). As it is seen from Fig.3, the spectral density of the radiation (transition radiation) is a monotonic function of the frequency and it is considerably smaller compared with the spectral density of CR in Fig.2.

## 4. Conclusions

In this paper we have derived analytical expressions for the spectral and spectral-angular distributions of the radiation energy from a charged particle flying through the centre of a dielectric ball. Numerical results for the spectrum of the radiation energy from an electron with the energy 2 MeV are given for two cases when the Cherenkov condition (17) is satisfied (Fig.2) and when it is not satisfied (Fig.3). The dispersion of the electromagnetic waves inside the ball material is not taken into account in numerical calculations. According to the data given in Figs.2,3, the spectral density of the radiated energy in the first case is larger in a few orders of magnitude than the corresponding quantity in the second one. One can explain this difference by the dominance of CR over the transition radiation in the frequency range under consideration.

In addition to the overall increase of the radiation intensity, narrow peaks appear in Fig.2 at certain "resonant" frequencies. These peaks are related with the influence of the ball-vacuum interface on the CR generated by the particle inside the ball. It is noteworthy, that the total energy of the radiation (in the range of frequencies under consideration) is approximately equal to the radiation energy from an electron moving rectilinearly in an infinite transparent medium with $\varepsilon_0 = \varepsilon_0'$ when the radiation is accumulated from the path with the length equal to the diameter of the ball. This fact indicates that the narrow peaks at "resonant" frequencies in Fig.2 are due to the spectral redistribution of the radiation influenced by the multiple reflection of CR from the ball-vacuum interface.

**Acknowledgements**

The author is grateful to L.SH. Grigoryan, A.A. Saharian, H.F. Khachatryan and A.S. Kotanjyan for stimulating discussions and continued interest to this work.